\documentclass{article}

\usepackage{arxiv}

\usepackage[utf8]{inputenc} 
\usepackage[T1]{fontenc}    
\usepackage{hyperref}       
\usepackage{url}            
\usepackage{booktabs}       
\usepackage{amsfonts}       
\usepackage{nicefrac}       
\usepackage{microtype}      
\usepackage{lipsum}
\usepackage{graphicx}
\usepackage{physics}
\usepackage{amsmath}
\graphicspath{ {./images/} }

\title{Effects of time dependence in the second order coherence of pulsed light}

\author{
 Joscelyn van der Veen \\
  Department of Physics\\
  University of Toronto\\
  Toronto, Canada \\
  \texttt{joscelyn.vanderveen@utoronto.ca} \\
   \And
 Daniel James \\
  Department of Physics\\
  University of Toronto\\
  Toronto, Canada \\
}

\begin{document}
\maketitle
\begin{abstract}
For light in a single temporal mode, the time dependence is associated entirely with the deterministic pulse shape rather than the quantum state. This means the second order coherence of light can no longer be understood as the temporal distribution of photons. We calculate the distribution of photon arrival time differences for pulsed light and find that it is inherently strongly bunched in terms of photocount arrival time differences. We further show how to account for the pulse shape when determining the second order coherence that characterizes quantum states. 
\end{abstract}

\begin{figure}[h]
    \centering
    \includegraphics[width=0.9\linewidth]{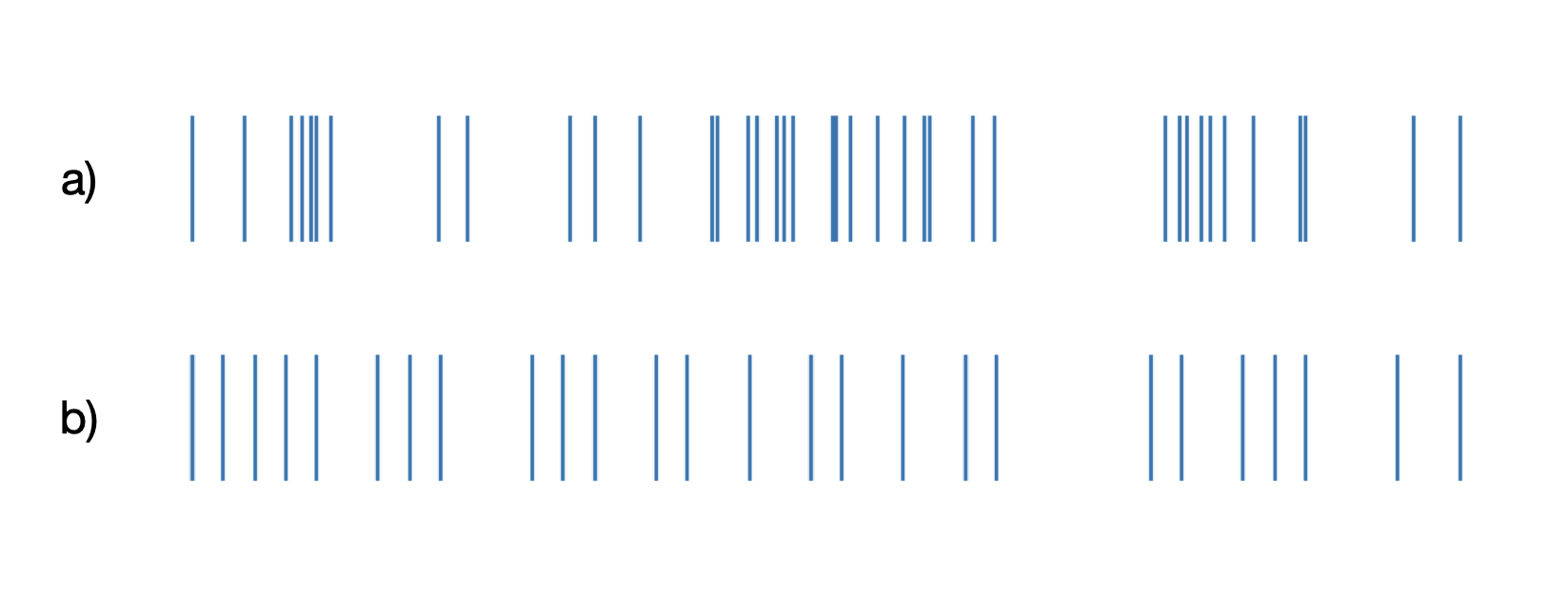}
    \caption{Example of photon counts as a function of time for a) thermal light with a Gaussian distribution and b) coherent light with a Poissonian distribution}
    \label{fig:photocounts}
\end{figure}

Photon bunching is the phenomena of photodetector counts clustering together in time more than can be expected from independently (Poissonian) distributed photocounts (Fig. \ref{fig:photocounts}). The measurement of photocount time distributions at a single photon detector \cite{Arecchi1966} gave an intuitive interpretation to the unintuitive intensity correlation results of Hanbury Brown and Twiss \cite{Brown1956,Loudon2000}. We can associate these two different measurements because they both depend on the second order coherence function $g^{(2)}(\tau)$, a function that is very important for the development of quantum technologies due to its use for quantum state identification \cite{Al-Qasimi_James_2025,CryerJenkins2023,Hassler2015}. Quantum state identification is necessary because the probabilistic nature of quantum mechanics means that we cannot measure a state and can only deduce a likely state from many measurements \cite{Smithey1993,James2001}. The values $g^{(2)}(0)=2$, $g^{(2)}(0)<1$, and $g^{(2)}(0)>2$ are called photon bunching, antibunching, and superbunching (or super-chaotic) respectively \cite{Loudon2000} because of the association between photocount time distributions and the second order coherence function. 

The second order coherence function is well-defined mathematically at some fixed position as \cite{Gerry2004},

\begin{equation}
    g^{(2)}(t,\tau)=\frac{\Tr{\rho E^{(-)}(t)E^{(-)}(t+\tau)E^{(+)}(t+\tau)E^{(+)}(t)}}{\Tr{\rho E^{(-)}(t)E^{(+)}(t)}\Tr{E^{(-)}(t+\tau)E^{(+)}(t+\tau)}}
    \label{eq:g2-tau}
\end{equation}

However, the exact correspondence between photodetector counts in time and the $g^{(2)}(\tau)$ function only exists for stationary light. The difficulty in determining $g^{(2)}(\tau)$ for pulsed light is clear from the different possible definitions \cite{Christ2011} and predictions \cite{Gorlach2020, stammer2025}. The primary source of this difficulty is the deterministic time dependence of the field, which introduces an inequivalence between the mathematical second order coherence function (Eq. \ref{eq:g2-tau}) used for theoretical quantum optics and the second order coherence function as a measure of photon bunching. In this letter, we show how this discrepancy in the second order coherence function arises for pulsed light, how this can result in misleading values of the second order coherence function, and for which applications the varying definitions are a useful measure. 

First, consider stationary light of intensity $I(t)$ falling on a detector. The probability of a detection in the interval $t$ to $t+\Delta t$ is $p(t)=s \expval{I}$, where $\expval{I}$ is the average photocount rate and $s$ is a parameter determined by the efficiency of the detector. This intensity will be measured as a series of detections in time and we could find a distribution of the time differences between adjacent counts. The distribution of photocounts is the conditional probability $p_c(t+\tau|t)$: the probability of counting a photon in the time interval $t+\tau$ to $t+\tau+\Delta\tau$ given that a photon has been counted in the time interval $t$ to $t+\Delta t$. For a stationary field, the conditional probability depends only on the time difference $\tau$ and is given by \cite[Eq. 6.22,]{Mandel1965},

\begin{equation}
    p_c(t+\tau|t)=s \frac{\expval{I(t)I(t+\tau)}}{\expval{I}}.
\end{equation}

\begin{figure}
    \centering
    \includegraphics[width=0.9\linewidth]{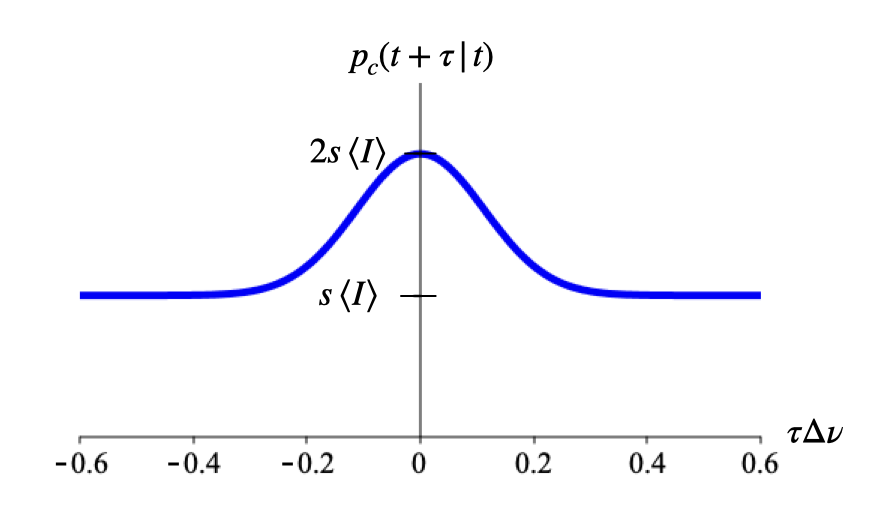}
    \caption{The conditional probability distribution for a stationary thermal source with detector efficiency $s$, average intensity $\expval{I}$, and a spectral bandwidth $\Delta\nu$.}
    \label{fig:stationary-pc}
\end{figure}

Fig. \ref{fig:stationary-pc} shows the conditional probability for stationary thermal (chaotic) light. The curve in the conditional probability for thermal light is known as a Hanbury Brown-Twiss (HBT) peak and its width is used to determine the coherence time of a thermal source \cite{Ferrantini2025}. The height of the HBT peak corresponds to $g^{(2)}(0)=2$ when normalized. Importantly, this normalization is the value $s\expval{I}$, which is the rate at which a detector counts photons from the light source. With stationary light, if we measure the conditional probability for a longer period of time then the value of the conditional probability will increase proportionally for all $\tau$. As we will see, this is not the case for a pulsed light source.

Consider now the conditional probability $p_c(t+\tau|t)$ for a pulsed light source. By definition, the conditional probablility of measuring a photocount in the time interval around $t+\tau$ can be written as the quotient of the probability of measuring a photocount around both $t$ and $t+\tau$, $p(t\cap t+\tau)$, and the probability of measuring a photocount around $t$, $p(t)$, so

\begin{equation}
    p_c(t+\tau|t)=\frac{p(t\cap t+\tau)}{p(t)}.
    \label{eq:cond-prob}
\end{equation}

The probability of measuring $n$ counts in the time intervals $\{t_{0,1},\ldots,t_{0,n}\}$ to $\{t_1,\ldots,t_n\}$ is given \cite[Eq.  2.68,]{Glauber2006}

\begin{multline}
    p^{(n)}(t_1,\ldots,t_n;t_{0,1},\ldots,t_{0,n})=s^n\int_{t_{0,1}}^{t_1}\dd{t_1'}\ldots\int_{t_{0,n}}^{t_n}\dd{t_n'} \\
    G^{(n)}(\boldsymbol{r}_1,t_1';\ldots;\boldsymbol{r}_n,t_n';\boldsymbol{r}_n,t_n';\ldots;\boldsymbol{r}_1,t_1')
    \label{eq:pn}
\end{multline}

\noindent where the correlation function $G^{(n)}(\boldsymbol{r}_1,t_1;\ldots;\boldsymbol{r}_{2n},t_{2n})$ is defined as,

\begin{multline}
    G^{(n)}(\boldsymbol{r}_1,t_1;\ldots;\boldsymbol{r}_{2n},t_{2n})\equiv\Tr\left\{\rho\boldsymbol{E}^{(-)}(\boldsymbol{r}_1,t_1)\ldots\right. \\
    \left.\times\boldsymbol{E}^{(-)}(\boldsymbol{r}_n,t_n)\boldsymbol{E}^{(+)}(\boldsymbol{r}_{n+1},t_{n+1})\ldots\boldsymbol{E}^{(+)}(\boldsymbol{r}_{2n},t_{2n})\right\}.
    \label{eq:Gn}
\end{multline}

It is important to note here that the electric field operators $\boldsymbol{E}^{(\pm)}(\boldsymbol{r},t)$ are measurement operators. Any pulse shape of the light is independent of the measurement operators and must be contained in the density matrix, $\rho$, of the field we are measuring. Since the density matrix describes the initial state of the field, it is not time dependent and contains the pulse shape through a polychromatic frequency distribution. 

While the electric field operators are typically written as sums of monochromatic waves, we can always make a change of basis. In a spatial-temporal mode basis $\{\hat{b}_j\}$, the positive frequency electric field operator can be written as \cite{Titulaer1966},

\begin{equation}
    \boldsymbol{E}^{(+)}(\boldsymbol{r},t)=\sum_j\hat{b}_j\boldsymbol{v}_j(\boldsymbol{r},t)
    \label{eq:E-temp-basis}
\end{equation}

\noindent where $\boldsymbol{v}_j(\boldsymbol{r},t)$ are polychromatic solutions of the wave equation.

If we have light traveling in a beam-like geometry, like a laser, then we can consider a single temporal mode $\hat{b}$ whose corresponding polychromatic solution $v(t)$ is the deterministic time dependence of the light. For example, the common basis choice of Hermite-Gauss polynomials \cite{Raymer2020} has the $j=0$ temporal mode as a Gaussian, which can describe a laser pulse of width $\Delta t_p$ as the temporal mode shape $v_0(t)=\exp(-t^2/2\Delta t_p^2)$. Unless we go into the subcycle pulse regime, we can choose $v_0(t)$ to be the temporal pulse shape because we can always generate a set of orthonormal basis functions that include the Fourier transform of the temporal pulse shape.

Let us assume the light is beam-like and in a single temporal mode $\hat{b}$. As with continuous light, we can approximate light as single mode if detection occurs on a timescale that is short compared to the coherence time, where coherence time refers to the phase diffusion due to random (non-quantum state) fluctuations \cite{Glauber2006,Mandel_Wolf_1995}. For pulsed light, this requires pulses whose duration are short compared to any statistical fluctuations that reduce interference effects. Then with Eqs. \ref{eq:pn}, \ref{eq:Gn}, and \ref{eq:E-temp-basis} we can write the conditional probability of a count in the time interval $\Delta \tau$ as,

\begin{align}
\begin{split}
    p_c(t+\tau|t)\Delta\tau&=\frac{p^{(2)}(t+\Delta t,t+\tau+\Delta\tau;t,t+\tau)}{p^{(1)}(t+\Delta t,t)} \\
    &=s\frac{\Tr{\rho\left(\hat{b}^\dagger\right)^2\hat{b}^2}}{\Tr{\rho\hat{b}^\dagger\hat{b}}}\int_{t+\tau}^{t+\tau+\Delta\tau}\dd{t''}\left|v(t'')\right|^2.
\end{split}
\end{align}

Since $\Delta\tau$ is a small time period, we thus find that in a single temporal mode, the probability of measuring a count in the time interval $t+\tau$ to $t+\tau+\Delta\tau$ given there was a count in the time interval $t$ to $t+\Delta t$ is,

\begin{equation}
    p_c(t+\tau|t)=s\frac{\Tr{\rho\left(\hat{b}^\dagger\right)^2\hat{b}^2}}{\Tr{\rho\hat{b}^\dagger\hat{b}}}\left|v(t+\tau)\right|^2.
    \label{eq:pc-temp-mode}
\end{equation}

The conditional probability clearly now depends on $t$ as well as $\tau$ as we would expect for the nonstationary pulsed light. However, we can experimentally measure a distribution of time differences between photocounts even for pulsed light, and that is a function only of the time difference $\tau$. To get this photon time difference distribution, which we call $D(\tau)$, we must average the conditional probability over the arrival time of the first photocount $t$ for each of the $N$ pulses we measure. As we have already seen, the probability of measuring a count in the time interval $t$ to $t+\Delta t$ is given by $p^{(1)}(t+\Delta t;t)$, which is proportional to the infinitesimal time interval $\Delta t$. We can thus obtain $D(\tau)$ by integrating over the infinitesimal $\Delta t$ and multiplying by the number of pulses $N$,

\begin{equation}
    D(\tau)=Ns^2\Tr{\rho\left(\hat{b}^\dagger\right)^2\hat{b}^2}\int_{-\infty}^\infty\dd{t}\left|v(t+\tau)\right|^2\left|v(t)\right|^2.
\end{equation}

The shape of $D(\tau)$ is the intensity autocorrelation function of the pulse shape. This is interesting because intensity autocorrelation measurements typically requires a nonlinear optical process \cite{Paschotta2007}. Fig. \ref{fig:I-tau} shows $D(\tau)$ for a Gaussian temporal mode of width $\Delta t_p$. Unlike the stationary case (Fig. \ref{fig:stationary-pc}) where the conditional probability was Gaussian due to interference of chaotic multimode light, the Gaussian curve due to deterministic pulse shape has no normalization for the height of the peak of $D(\tau)$ within the distribution itself. We also no longer have a stationary average intensity $\expval{I}$ but we can measure the intensity of $N$ pulses:

\begin{figure}
    \centering
    \includegraphics[width=0.9\linewidth]{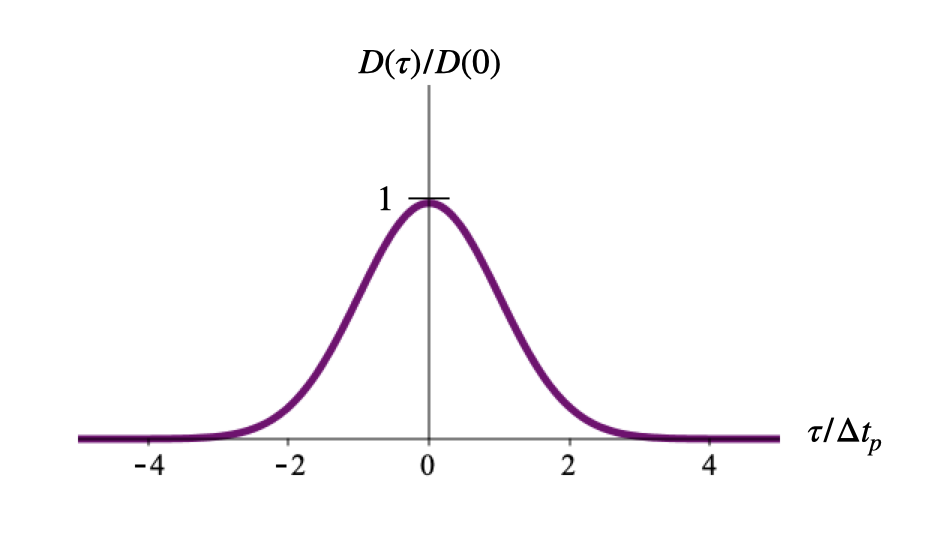}
    \caption{Time difference distribution $D(\tau)$ assuming a Gaussian temporal mode $v(t)=\exp(-t^2/2\Delta t_p^2)$. Unlike the distribution of stationary photocounts, the photocounts with long time differences go to zero beyond the time of the pulse and therefore the long time behaviour of $D(\tau)$ cannot be used for normalization.}
    \label{fig:I-tau}
\end{figure}

\begin{equation}
    I_p(N)=Ns\Tr{\rho\hat{b}^\dagger\hat{b}}\int_{-\infty}^\infty\dd{t'}\left|v(t')\right|^2
\end{equation}

\noindent which now depends on the number of pulses measured and the temporal mode function. 

The $N$-pulse intensity has equivalent units to the average intensity over a measurement integration time $\tau_{M}$, $\expval{I}\tau_M$, because $I_p(N)$ is an integrated intensity rather than a rate.

Writing the time difference distribution in terms of the $N$-pulse intensity yields,

\begin{equation}
    D(\tau)=I_p(N)^2g^{(2)}_q\frac{\eta(\tau)}{N}
\end{equation}

\noindent where we define 

\begin{align}
    g^{(2)}_q&\equiv\frac{\Tr{\rho\left(\hat{b}^\dagger\right)^2\hat{b}^2}}{\Tr{\rho\hat{b}^\dagger\hat{b}}^2} \\
    \eta(\tau)&\equiv\frac{\int_{-\infty}^\infty\dd{t}\left|v(t+\tau)\right|^2\left|v(t)\right|^2}{\left(\int_{-\infty}^\infty\dd{t'}\left|v(t')\right|^2\right)^2}.
\end{align}

The factor $g^{(2)}_q$ we call such because it depends only on the quantum state and it is equal to the $g^{(2)}(t,\tau)$ from Eq. \ref{eq:g2-tau} for a single mode field. This is the second order coherence function that we defined mathematically and it is the quantity that can identify the quantum state of the light.

However, for stationary light, the $g^{(2)}(\tau)$ originated as a measure of the physical correlations of photodetector counts in time with the $g^{(2)}(0)$ being a measure that identifies an amount of photon bunching. For pulsed light, we would measure the likelihood of photocounts occurring closely in time as,

\begin{equation}
    g^{(2)}_p\equiv D(0)/I_p(N)^2=g^{(2)}_q\frac{\eta(0)}{N}\neq g^{(2)}_q
\end{equation}

This difference between the mathematical second order coherence function $g^{(2)}_q$ and the tendency to measure photocounts closer or farther apart in time $g^{(2)}_p$ is the main result of this letter. 

Moreover, because there is no longer a singular definition for second order coherence, various experimental results will choose a normalization that yields a $g^{(2)}$-like quantity other than the $g^{(2)}_q$ or $g^{(2)}_p$. For example, some experimental results may take the temporal mode function $v(t)$ to extend over multiple pulses in the train and then normalize the central peak with either the integrated time differences \cite{Stevens2014} or the height \cite{Theidel2024} of the other peaks. In both these case, the reliance of the normalization on a portion of the temporal mode function means they observe a $g^{(2)}$ that increases with the power. If this were the stationary $g^{(2)}_q$ changing, it would imply the quantum state changes with power. It is therefore misleading when we do not specify that it is not the $g^{(2)}_q$ being measured due to the normalization. 

If we have multi-temporal mode light, there is potential for normalization from a comparison of photocounts originating from pulses that are separated by more than the coherence time of the pulse train. This requires a description of pulse to pulse correlations, which are not currently well understood, and is the topic of future investigations. 

The stationary $g^{(2)}_q$ is the quantity we should measure for information about the quantum state of light. We can determine $g^{(2)}_q$ from the time difference distribution $D(\tau)$ if we know the temporal mode function to calculation $\eta(\tau)$. For example, consider a Gaussian temporal mode function of temporal width $\Delta t_p$. Then,

\begin{equation}
    \eta(\tau,\Delta t_p)=\frac{1}{\sqrt{2\pi}\Delta t_p}\exp\left(\frac{-\tau^2}{2\Delta t_p^2}\right).
\end{equation}

Since $\eta(0)$ depends only on the pulse width $\Delta t_p$ and we can determine the pulse width from the time difference distribution $D(\tau)$, we could calculate $g^{(2)}_q$ as,

\begin{equation}
    g^{(2)}_q=\sqrt{2\pi}\Delta t_pN\frac{D(0)}{I_p(N)^2}
\end{equation}

This is an indirect calculation that has more sources of uncertainty than other measurements of the $g^{(2)}_q$ such as a photon number distribution measurement,

\begin{equation}
    g^{(2)}_q=\sum_nn(n-1)P_n/\bar{n}^2.
\end{equation}

\noindent where $P_n=\mel{n}{\rho}{n}$ is the photon number distribution of the single mode field defined by the density matrix $\rho$.

\begin{figure}
    \centering
    \includegraphics[width=0.9\linewidth]{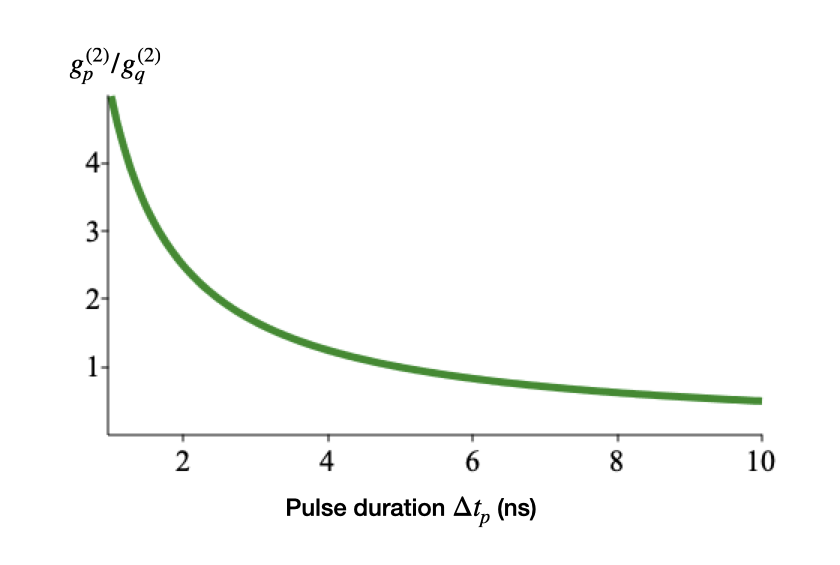}
    \caption{The pulsed $g^{(2)}_p$ for an arbitrary quantum state (normalized by $g^{(2)}_q$) for Gaussian pulses with width $\Delta t_p$ and an 80MHz pulse repetition rate.}
    \label{fig:I-per-s}
\end{figure}

The pulsed $g^{(2)}_p$ does not yield information about the quantum state of the light but it is the best measure of photon correlations. Since photocounts can only be measured during pulses, pulses do cause a bunching of photocounts in time. This bunching is due to the deterministic pulse shape rather than the statistics of light but it can still be used in applications that require a measure of correlations in time, such as time-bin encoding \cite{Fenwick2024}. From Fig. \ref{fig:I-per-s}, we can see that decreasing the pulse duration will increase $g^{(2)}_p$ by bunching the photocounts within a smaller pulse duration. 

When measuring a distribution of time differences, we can thus extract a variety of second order coherence-like functions depending on our choice of normalization. The second order coherence function that is germane to quantum technologies is the $g^{(2)}_q$ that identifies the quantum state of light, and it can only be extracted through normalization that includes information such as the pulse duration. The second order coherence instead obtained just through normalization by the pulse intensity, the $g^{(2)}_p$, is still useful for describing all photon correlations that exist in pulsed light, both from the quantum state and the deterministic pulse shape.

\bibliographystyle{unsrt}  
\bibliography{references}  


\end{document}